\documentclass[prd,aps,showpacs,nofootinbib,amsmath]{revtex4}
\usepackage{epsf,epsfig,graphicx}

\begin{document}

\title{Three-parton contribution to pion form factor in $k_T$ factorization}

\author{Yu-Chun Chen}
\email{ycchen@phys.sinica.edu.tw}
\author{Hsiang-nan Li}
\email{hnli@phys.sinica.edu.tw}

\affiliation{Institute of Physics, Academia Sinica, Taipei, Taiwan
115, Republic of China} \affiliation{Department of Physics,
Tsing-Hua University, Hsinchu, Taiwan 300, Republic of China}
\affiliation{Department of Physics, National Cheng-Kung University,\\
Tainan, Taiwan 701, Republic of China}

\begin{abstract}

We set up a framework for the study of the power-suppressed
three-parton contribution to the pion electromagnetic form factor in
the $k_T$ factorization theorem. It is first shown that the gauge
dependence proportional to parton transverse momenta from the
two-parton Fock state and the gauge dependence associated with the
three-parton Fock state cancel each other. After verifying the gauge
invariance, we derive the three-parton-to-three-parton
$k_T$-dependent hard kernel at leading order of the coupling
constant, and find that it leads to about 5\% correction to the pion
electromagnetic form factor in the whole range of experimentally
accessible momentum transfer squared. This subleading contribution
is much smaller than the leading-order twist-2,
next-to-leading-order twist-2 and leading-order two-parton twist-3
ones, which have been calculated in the literature.

\end{abstract}


\maketitle

\section{INTRODUCTION}

Aspects of the $k_T$ factorization theorem
\cite{CCH,CE,LRS,BS,LS,HS} in perturbative QCD have been
investigated intensively. One of the important issues is about the
derivation of a $k_T$-dependent hard kernel at subleading level,
which is defined as the difference between QCD diagrams and
effective diagrams for transverse-momentum-dependent (TMD) hadron
wave functions. We have explained that partons in both sets of
diagrams should remain off mass shell by $k_T^2$ in the $k_T$
factorization theorem \cite{NL2}. The same statement has been made
in the application of the $k_T$ factorization theorem to inclusive
processes such as prompt photon production \cite{LMZ11}. The
off-shellness of partons may cause concern of the gauge invariance
\cite{FMW08,FMW11}\footnote{Criticisms raised in \cite{FMW08,FMW11}
have been responded in \cite{LM09,NL09}.}. However, we have shown
that the gauge dependence cancels between the above two sets of
diagrams, and a $k_T$-dependent hard kernel is gauge invariant
\cite{LM09,NL07}. Following this prescription, the
next-to-leading-order (NLO) correction to the pion transition
(electromagnetic) form factor associated with the process
$\pi\gamma^*\to \gamma(\pi)$ has been calculated at leading twist,
i.e., twist 2 \cite{NL07,LWS10}. Here we shall study the
power-suppressed three-parton contribution to the pion
electromagnetic form factor in the $k_T$ factorization theorem. The
three-parton contribution in the collinear factorization theorem
\cite{LB80} to a simpler process, the $\rho$ meson transition form
factor, has been evaluated recently \cite{AIP08}.

We shall first demonstrate the gauge invariance of the three-parton
contribution to the pion electromagnetic form factor in the $k_T$
factorization theorem. There are two sources of gauge dependence for
this power correction \cite{Q90}: the first source is proportional
to parton transverse momenta from the two-parton Fock state. The
corresponding hadronic matrix element is written as
\begin{eqnarray} \langle
0|{\bar q}(z)\Gamma i\partial_\alpha q(0) |\pi\rangle, \label{d3}
\end{eqnarray}
where $z$ is the coordinate of the anti-quark field $\bar q$,
and $\Gamma$ represents a combination of Gamma matrices.
The second source is associated with the three-parton Fock state with an
additional valence gluon. The corresponding matrix element is given
by
\begin{eqnarray}
\langle 0|{\bar q}(z)\Gamma
%\sigma^{+\alpha'}\gamma_5
gT^aA_\alpha^a(z') q(0) |\pi\rangle, \label{a3}
\end{eqnarray}
with a color matrix $T^a$, and the gluon field $A_\alpha^a$ at the
coordinate $z'$. The gauge dependences from the above two sources
cancel each other, when Eqs.~(\ref{d3}) and (\ref{a3}) are combined
to form the gauge-invariant matrix element
\begin{eqnarray}
\langle 0|{\bar q}(z)\Gamma iD_\alpha(z') q(0)
|\pi\rangle,\label{t3}
\end{eqnarray}
with the covariant derivative $iD_\alpha \equiv
i\partial_\alpha+gT^aA_\alpha^a$. The cancellation of the gauge
dependences is similar to that occurring in the collinear
factorization theorem \cite{Q90}. In the $k_T$ factorization theorem
we just keep the transverse momentum dependence in denominators of
particle propagators \cite{NL2}. Hence, it is natural that the gauge
dependence disappears at higher twists in the same way as in the
collinear factorization theorem.

Our formalism implies that contributions proportional to transverse
momenta in numerators of hard kernels must be combined with
contributions from three-parton Fock states in order to guarantee
gauge invariance. Therefore, there is concern on the study of the
pion transition form factor in \cite{CHM96}, where only the former
was included. It has been pointed out explicitly that the
leading-order (LO) hard kernel for the pion electromagnetic form
factor becomes gauge-dependent, if one simply considers parton
transverse momenta in numerators \cite{FMW11}. However, the
contribution from the three-parton Fock state was still missing in
\cite{FMW11}, such that the false postulation on the gauge
dependence of the $k_T$ factorization theorem was made.

Since both the initial- and final-state pions involve higher-twist matrix
elements like that in Eq.~(\ref{t3}), the three-parton contribution
to the pion electromagnetic form factor is suppressed at least by
$1/Q^2$, $Q^2$ being the momentum transfer squared. After examining
the gauge invariance, we calculate the $k_T$-dependent hard kernel
for the three-parton-to-three-parton scattering in the Feynman
gauge, and convolute it with the three-parton pion wave functions.
We observe that the diagrams with a four-gluon vertex dominate this
power correction. It will be shown that the three-parton
contribution is only about 5\% of the sum of those which have been
investigated before, including the LO twist-2, NLO twist-2, and LO
two-parton twist-3 ones \cite{LWS10}. That is, the three-parton
contribution is not crucial for accommodating experimental data of
the pion electromagnetic form factor. At the same power of $1/Q^2$,
one should also take into account the scattering of two (four)
partons into four (two) partons in principle. This piece has been
analyzed in light-cone sum rules \cite{BK02}, and found to be less
important than other contributions. With this work, we conclude that
the chirally enhanced two-parton twist-3 correction is the most
important $1/Q^2$ correction to the pion electromagnetic form
factor.

In Sec.~II we verify the gauge invariance of the LO three-parton
contribution to the pion electromagnetic form factor by combining
the gauge-dependent hard kernels corresponding to Eqs.~(\ref{d3})
and (\ref{a3}). The $k_T$-dependent hard kernel is then derived from
the three-parton-to-three-parton scattering diagrams, and convoluted
with the three-parton pion wave functions numerically in Sec.~III.
Section IV is the conclusion. Detailed calculations of the
gauge-dependent hard kernels corresponding to Eq.~(\ref{a3}) are
presented in Appendix A, and the expressions of the
three-parton-to-three-parton hard kernels are collected in Appendix
B.

\section{GAUGE INVARIANCE}

\begin{figure}[t]
\begin{center}
\includegraphics[height=3.3cm]{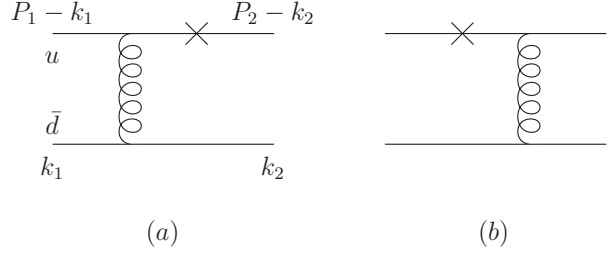}
\caption{LO diagrams for the pion electromagnetic form factor, where
the symbol $\times$ represents the virtual photon vertex.}
\label{fig1} \vspace{-0.5cm}
\end{center}
\end{figure}

Consider the pion electromagnetic form factor involved in the
process $\pi(P_1)\gamma^*\to\pi(P_2)$, whose LO diagrams are
displayed in Fig.~\ref{fig1}. The momentum $P_1$ of the
initial-state pion and $P_2$ of the final-state pion are
parameterized as
\begin{eqnarray}
P_1=(P_1^+,0,{\bf 0}_T)=\frac{Q}{\sqrt{2}}(1,0,{\bf 0}_T),\;\;\;\;
P_2=(0,P_2^-,{\bf 0}_T)=\frac{Q}{\sqrt{2}}(0,1,{\bf 0}_T),
\label{mpp}
\end{eqnarray}
with $Q^2=-q^2$, $q=P_2-P_1$ being the virtual photon momentum. The
gluon propagator of momentum $l$ is written as
\begin{eqnarray}
\frac{-i}{l^2}\left(g^{\sigma\nu} - \lambda \frac{l^\sigma
l^\nu}{l^2}\right),\label{tensor}
\end{eqnarray}
in the covariant gauge, where the parameter $\lambda$ will be used
to identify sources of gauge dependence. We assume that the
anti-quarks in the initial- and final-state pions, represented by
lower fermion line, carry the parton momenta
\begin{eqnarray}
k_1=(x_1P_1^+,0,{\bf k}_{1T}),\;\;\;\; k_2=(0,x_2P_2^-,{\bf
k}_{2T}),
\end{eqnarray}
respectively, $x_1$ and $x_2$ being the momentum fractions. It is
understood that the components $k_1^-$ and $k_2^+$ have been dropped
in hard kernels, and integrated out of the TMD pion wave functions.

We employ the Fierz identity
\begin{eqnarray}
I_{ij}I_{lk}=\frac{1}{4}I_{ik}I_{lj}
+\frac{1}{4}(\gamma_5)_{ik}(\gamma_5)_{lj}
+\frac{1}{4}(\gamma^\alpha)_{ik}(\gamma_\alpha)_{lj}
+\frac{1}{4}(\gamma_5\gamma^\alpha)_{ik}(\gamma_\alpha\gamma_5)_{lj}
+\frac{1}{8}(\sigma^{\alpha\beta}\gamma_5)_{ik}
(\sigma_{\alpha\beta}\gamma_5)_{lj},\label{fie}
\end{eqnarray}
to factorize the fermion flow, where $I$ denotes the $4\times 4$
identity matrix. The structure $\gamma_\alpha\gamma_5$ in the above
identity contributes at twist 2 and higher twists, and $\gamma_5$
and $\sigma_{\alpha\beta}\gamma_5$ contribute at twist 3 and higher
twists at two-parton level. The matrix $\Gamma$ in Eqs.~(\ref{d3})
and (\ref{a3}) can pick up one of the above structures, among which
we focus on the first one $\gamma_{\alpha}\gamma_5$ as an example
below. We also insert the identity
\begin{eqnarray}
I_{ij}I_{lk}=\frac{1}{N_c}I_{lj}I_{ik}+2
(T^c)_{lj}(T^c)_{ik},\label{col}
\end{eqnarray}
to factorize the color flow, where $N_c=3$ is the number of colors,
$I$ denotes the $3\times 3$ identity matrix, and $T^c$ is a color
matrix. The first (second) term in Eq.~(\ref{col}) will be
associated with a color-singlet (color-octet) state of the valence
quark and anti-quark.

The first source of gauge dependence is extracted from the diagrams
in Fig.~\ref{fig1}, where the quark and anti-quark pair forms a
color-singlet state. Combining the decompositions in
Eqs.~(\ref{fie}) and (\ref{col}), we sandwich Fig.~\ref{fig1} with
the structures
\begin{eqnarray}
\frac{1}{4N_c}\gamma_{\alpha}\gamma_5,\;\;\;\;
\frac{1}{4N_c}\gamma_5\gamma_{\beta},\label{str}
\end{eqnarray}
from the initial and final states, respectively, where the
subscripts $\alpha$ and $\beta$ can take arbitrary components. The
LO hard kernel from Fig.~\ref{fig1}(a) contains the gauge-dependent
piece
\begin{eqnarray}
H^{a\lambda}=-ieg^2\lambda \frac{C_F}{16N_c}\frac{tr[ \gamma^\sigma
\gamma_5\gamma_{\beta}\gamma_\mu (\not P_1-\not
k_2)\gamma^\nu\gamma_{\alpha}\gamma_5]}
{(P_1-k_2)^2(k_1-k_2)^2}\frac{(k_1-k_2)_\sigma
(k_1-k_2)_\nu}{(k_1-k_2)^2},\label{hlam}
\end{eqnarray}
with $C_F$ being a color factor. The above expression diminishes
with the substitution $k_1= x_1P_1$, $k_2= x_2P_2$,
$\gamma_\alpha=\gamma^-$ (proportional to $\not\! P_1$), and
$\gamma_\beta=\gamma^+$ (proportional to $\not\! P_2$) into the
numerator, implying that Eq.~(\ref{hlam}) does not contribute at
leading power in the $k_T$ factorization theorem.

To obtain the gauge-dependent hard kernel from Eq.~(\ref{hlam}) at
$1/Q^2$, we insert the identity
$(k_1-k_2)_\nu=(P_1-k_2)_\nu-(P_1-k_1)_\nu$. It can be shown that
the contribution from the $(P_1-k_2)_\nu$ term is cancelled by the
corresponding one in Fig.~\ref{fig1}(b). The second term, with
$P_1-k_1$ being the momentum of the incoming valence quark,
corresponds to the matrix element with the derivative of the quark
field in the initial-state pion. This term can be picked up by
differentiating Eq.~(\ref{hlam}) with respect to $k_1$. Once
Eq.~(\ref{d3}) for the initial-state pion is identified, we
pick up the $k_{2\sigma}$ term in $(k_1-k_2)_\sigma$ via 
differentiation, which
corresponds to the derivative of the valence anti-quark field in the
final-state pion. Note that denominators of particle propagators,
depending on $k_1$ and $k_2$, will be differentiated too. However,
their differentiation gives rise to even higher-twist matrix elements,
and can be neglected. We then extract
\begin{eqnarray}
H^{a\lambda}_{TT}(k_1,k_2)\equiv\frac{\partial^2 H^{a\lambda}}{\partial
k_{1\alpha}\partial k_{2\beta}} &=&ieg^2\lambda
\frac{C_F}{16N_c}\frac{tr[ \gamma^\beta \gamma_5\gamma_{\beta}\gamma_\mu
(\not P_1-x_2\not P_2)\gamma^\alpha\gamma_{\alpha}\gamma_5]}
{(P_1-k_2)^2(k_1-k_2)^4},\label{tt}
\end{eqnarray}
associated with $\langle 0|{\bar q}(z)\gamma_5\gamma^{\alpha}
i\partial_\alpha q(0) |\pi(P_1)\rangle$ and the similar matrix
element for the final-state pion.

The LO hard kernel from Fig.~\ref{fig1}(b) contains
\begin{eqnarray}
H^{b\lambda}=-ieg^2 \lambda\frac{C_F}{16N_c}\frac{tr[ \gamma^\sigma
\gamma_5\gamma_{\beta}\gamma^\nu(\not P_2-\not
k_1)\gamma_\mu \gamma_{\alpha}\gamma_5]}{(P_2-k_1)^2(k_1-k_2)^2}\frac{(k_1-k_2)_\sigma
(k_1-k_2)_\nu}{(k_1-k_2)^2},\label{hlbm}
\end{eqnarray}
whose differentiation with respect to $k_{1\alpha}$ and $k_{2\beta}$
leads to
\begin{eqnarray}
H^{b\lambda}_{TT}(k_1,k_2)&=&ieg^2\lambda \frac{C_F}{16N_c}\frac{tr[
\gamma^\alpha \gamma_5\gamma_{\beta}\gamma^\beta(\not
P_2-x_1\not P_1)\gamma_\mu\gamma_{\alpha}\gamma_5]}
{(P_2-k_1)^2(k_1-k_2)^4}.\label{ttb}
\end{eqnarray}
This hard kernel corresponds to the matrix element of the
initial-state pion with the derivative of the anti-quark field $\bar
q(z)$. Equations~(\ref{tt}) and (\ref{ttb}) represent the gauge
dependence in LO two-parton-to-two-parton scattering at power of
$1/Q^2$, which was also observed in \cite{FMW11}.

\begin{figure}[t]
\begin{center}
\includegraphics[height=2.8cm]{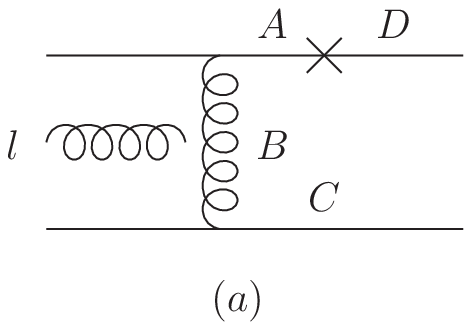}\hspace{1.0cm}
\includegraphics[height=2.8cm]{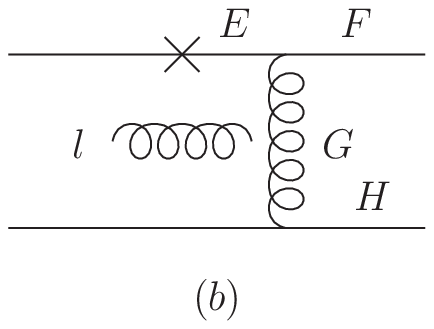}
\caption{Diagrams with three partons from the initial state, where
letters $A$, $B$, $\cdots$ denote the attachments of the additional
valence gluon.} \label{fig2}
\end{center}
\end{figure}

The second source of gauge dependence arises from Fig.~\ref{fig2},
where three partons appear in the initial state as indicated by
Eq.~(\ref{a3}). All possible attachments of the additional valence
gluon to the lines other than the valence quark and anti-quark in
the initial state are labeled by letters $A$, $B$, $\cdots$. In
principle, the diagrams with the attachments to the valence quark
and anti-quark in the initial state should be included in order to
respect $U(1)$ gauge symmetry for the electromagnetic
interaction\footnote{We thank V. Braun for pointing out this $U(1)$
gauge symmetry.}. These diagrams contribute to higher Gegenbauer
terms in the two-parton twist-3 pion distribution amplitudes.
Equations of motion can then be constructed to relate the
coefficients of the higher Gegenbauer terms in the two-parton
twist-3 and three-parton twist-3 pion distribution amplitudes
\cite{BF90}. Hence, one should pay attention to the consistency
between the models for these two sets of distribution amplitudes in
a numerical analysis. We shall adopt the non-asymptotic models for
both sets of distribution amplitudes in Sec.~III, when estimating
the importance of the three-parton contribution relative to other
two-parton contributions in the pion electromagnetic form factor.

According to Eq.~(\ref{col}), we absorb a color matrix $T^c$ and the
coupling constant $g$ associated with an attachment into the matrix
element for the initial-state pion, and another $T^c$ goes into the
evaluation of gauge-dependent hard kernels. For example, the color
factor corresponding to the attachment $A$ of the valence gluon to
the virtual quark line is given by
\begin{eqnarray}
tr[T^bT^aT^bT^c] =-\frac{1}{4N_c} \delta^{ac},\label{ca}
\end{eqnarray}
where the color matrix $T^a$ ($T^b$) comes from the valence (hard)
gluon vertex. After summing over $c$, the tensor $\delta^{ac}$ sets
$c=a$ in the matrix element for the initial-state pion, leading to
Eq.~(\ref{a3}). Including the coefficient 2 in Eq.~(\ref{col}), we
adopt the structure $\gamma_{\alpha}\gamma_5/2$ for the initial
state in the calculation of Fig.~\ref{fig2}, whose details can be
found in Appendix A. The results are collected as follows:
\begin{eqnarray}
H_{AT}^{\lambda}&=&-ieg^2\lambda\frac{1}{32N_c^2}
\frac{tr[\gamma^\beta \gamma_5\gamma_{\beta}\gamma_\mu (\not
P_1-x_2\not P_2)\gamma^\alpha\gamma_{\alpha}\gamma_5]}
{(P_1-k_2)^2(k_1-k_2)^4},\label{aA}\\
H_{BT}^{\lambda}
&=&ieg^2\lambda \frac{1}{32}\frac{tr[\gamma^\beta\gamma_5
\gamma_{\beta}\gamma_\mu (\not P_1-x_2\not P_2)\gamma^\alpha
\gamma_{\alpha}\gamma_5]}{(P_1-k_2)^2(k_1-k_2)^4},\label{B}\\
H^{\lambda}_{CT}&=&0,\label{C}\\
H^{\lambda}_{DT}&=&-ieg^2\lambda\frac{1}{32N_c^2}
\frac{tr[\gamma^\beta \gamma_5
\gamma_{\beta}\gamma^\alpha(\not P_2-x_2\not P_2-y_1\not
P_1)\gamma_\mu
\gamma_{\alpha}\gamma_5]}{(P_2-k_2-l_1)^2(k_1-k_2)^4},\label{D}\\
H^{\lambda}_{ET}&=&-ieg^2\lambda\frac{C_F}{16N_c}
\frac{tr[\gamma^\beta \gamma_5\gamma_{\beta} \gamma^\alpha (\not
P_2- x_1\not P_1-y_1\not P_1) \gamma_\mu
\gamma_{\alpha}\gamma_5]}{(P_2-k_1-l_1)^2(k_1-k_2)^4},\label{E}\\
H^{\lambda}_{FT}&=&ieg^2\lambda\frac{1}{32N_c^2}
\frac{tr[\gamma^\beta \gamma_5\gamma_{\beta}\gamma^\alpha(
\not P_2- x_2\not P_2-y_1\not P_1)\gamma_\mu
\gamma_{\alpha}\gamma_5]}
{(P_2-k_2-l_1)^2(k_1-k_2)^4}\nonumber\\
& &-ieg^2\lambda\frac{1}{32N_c^2} \frac{tr[\gamma^\beta
\gamma_5\gamma_{\beta} \gamma^\alpha(\not P_2-x_1\not P_1-y_1\not
P_1)\gamma_\mu
\gamma_{\alpha}\gamma_5]}{(P_2-k_1-l_1)^2(k_1-k_2)^4},\label{F}\\
H^{\lambda}_{GT} &=& ieg^2\lambda \frac{1}{32}\frac{tr[\gamma^\beta
\gamma_5\gamma_{\beta}\gamma^\alpha (\not P_2-x_1\not P_1-y_1\not
P_1)\gamma_\mu
\gamma_{\alpha}\gamma_5]}{(P_2-k_1-l_1)^2(k_1-k_2)^4}\nonumber\\
& &+ieg^2 \lambda\frac{1}{32}\frac{tr[\gamma^\alpha\gamma_5
\gamma_{\beta} \gamma^\beta(\not P_2-x_1\not P_1-y_1\not P_1)\gamma_\mu
\gamma_{\alpha}\gamma_5]}{(P_2-k_1-l_1)^2(k_1+l_1-k_2)^4},\label{G}\\
H^{\lambda}_{HT} &=&-ieg^2\lambda\frac{1}{32N_c^2} \frac{tr[
\gamma^\alpha\gamma_5\gamma_{\beta}\gamma^\beta(\not
P_2-x_1\not P_1-y_1\not P_1)\gamma_\mu
\gamma_{\alpha}\gamma_5]}{(P_2-k_1-l_1)^2 (k_1+l_1-k_2)^4},\label{H}
\end{eqnarray}
with the gluon momentum fraction $y_1=l_1^+/P_1^+$.

Summing the above expressions, we arrive at
\begin{eqnarray}
\sum_{i=A}^H
H_{iT}^\lambda=H^{a\lambda}_{gT}+H^{b\lambda}_{gT},
\end{eqnarray}
with
\begin{eqnarray}
H^{a\lambda}_{gT}=H^{a\lambda}_{TT}(k_1,k_2),\;\;\;\;
H^{b\lambda}_{gT}=H^{b\lambda}_{TT}(k_1+l_1,k_2).\label{aB}
\end{eqnarray}
The contributions from the attachments $A$ and $B$ are added into
$H^{a\lambda}_{gT}$ with the desired color factor $C_F$. The
contribution from the attachment $D$ and the first term in the
attachment $F$ cancels each other. The second term from the
attachment $F$ and the first term from the attachment $G$ are
combined into the expression with the color factor $C_F$, which then
cancels the contribution from the attachment $E$. The second term of
$G$ and the contribution from the attachment $H$ are added into
$H^{b\lambda}_{gT}$. Note that $H^{a\lambda}_{gT}$ does not depend
on the valence gluon momentum $l_1$, which can then be integrated
out of the matrix element, giving $\langle 0|{\bar
q}(z)\gamma_5\gamma^{\alpha} gT^aA_\alpha^{a}(0) q(0)
|\pi(P_1)\rangle$. The hard kernel $H^{b\lambda}_{gT}$, depending on
the combination $k_1+l_1$, corresponds to the matrix element
$\langle 0|{\bar q}(z)\gamma_5\gamma^{\alpha} gT^aA_\alpha^a(z)
q(0)|\pi(P_1)\rangle$. Because of the symmetry under the exchange of
the initial- and final-state kinematic variables, the
gauge-dependent hard kernels with three partons from the final state
are written as
\begin{eqnarray}
H^{a\lambda}_{Tg}=H^{a\lambda}_{TT}(k_1,k_2+l_2),\;\;\;\;
H^{b\lambda}_{Tg}=H^{b\lambda}_{TT}(k_1,k_2),\label{bB}
\end{eqnarray}
where $l_2$ is the momentum carried by the outgoing valence gluon.

\begin{figure}[t]
\begin{center}
\includegraphics[height=3.5cm]{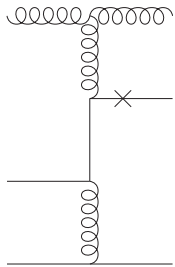}
\caption{One of the three-parton-to-three-parton diagrams, where the
two valence gluons scatter via a three-gluon vertex.} \label{figC}
\end{center}
\end{figure}

At last, we extract the gauge dependence from the
three-parton-to-three-parton diagrams.
In this case the indices $\alpha$ and $\beta$, associated with the
initial and final valence gluons, respectively, must be carried by
gluon vertices, instead of by parton momenta. Focusing on the
gauge-dependent piece, we can apply the Ward identity to hard
gluons. It is easy to find that Fig.~\ref{figC}, where the two
valence gluons scatter via a three-gluon vertex, does not
contribute: if the gauge dependence arises from the lower hard
gluon, the results, being proportional to the parton momenta after
applying the Ward identity, should be dropped. If the gauge
dependence arises from the upper hard gluon, the Ward identity
diminishes the amplitude for a similar reason. It is also easy to
see that Fig.~\ref{fig9}(a) with a four-gluon vertex does not
contribute to a gauge-dependent hard kernel. If both the valence
gluons attach to the quark line, the gauge-dependent contribution
vanishes because of the Ward identity applied to the anti-quark
line. If both the valence gluons attach to the anti-quark line, the
gauge-dependent contribution vanishes too.

\begin{figure}[t]
\begin{center}

\includegraphics[height=2.5cm]{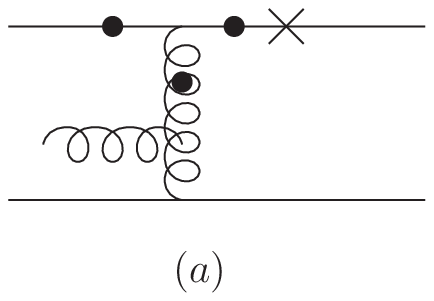}\hspace{0.5cm}
\includegraphics[height=2.5cm]{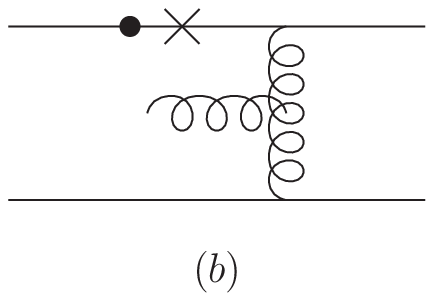}\hspace{0.5cm}
\includegraphics[height=2.5cm]{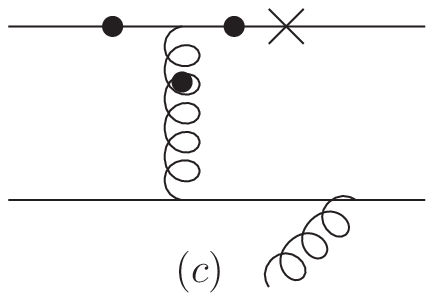}\hspace{0.5cm}
\includegraphics[height=2.5cm]{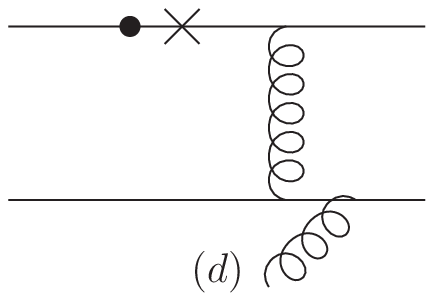}\vskip 0.5cm

\includegraphics[height=3.2cm]{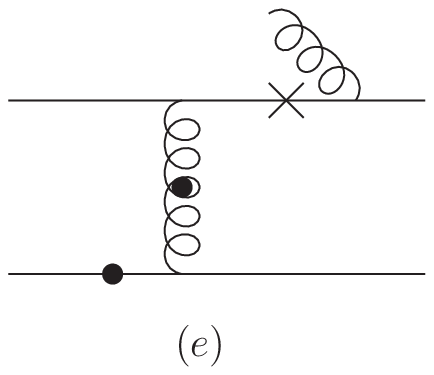}\hspace{0.5cm}
\includegraphics[height=3.2cm]{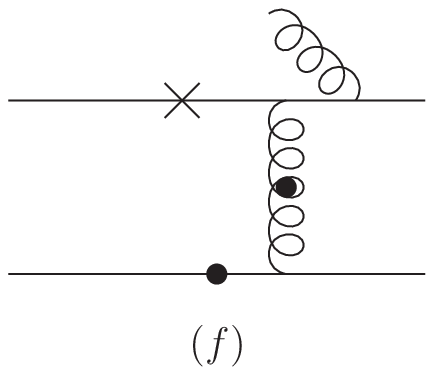}\hspace{0.5cm}
\includegraphics[height=3.2cm]{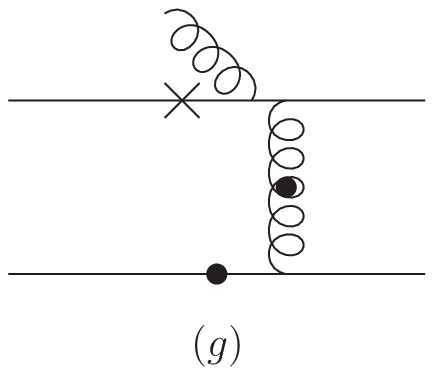}\hspace{0.5cm}
\includegraphics[height=2.5cm]{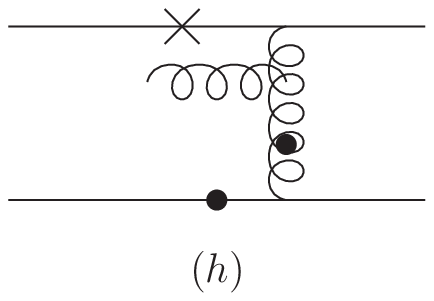}\vskip 0.5cm

\includegraphics[height=3.2cm]{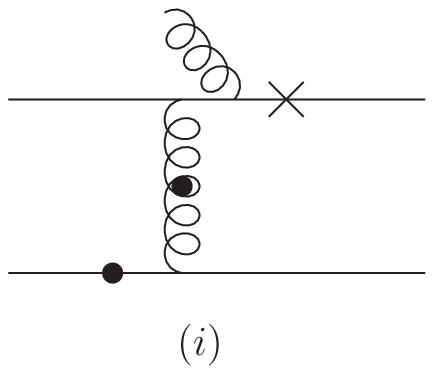}\hspace{0.5cm}
\includegraphics[height=2.5cm]{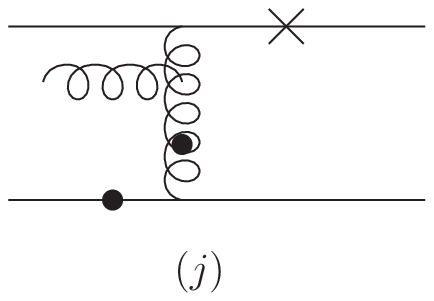}\hspace{0.5cm}
\includegraphics[height=2.5cm]{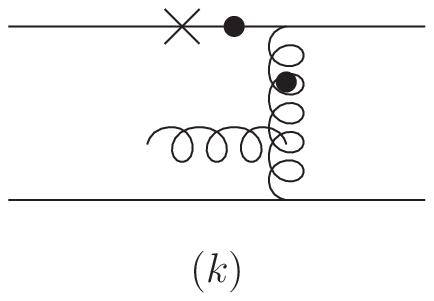}\hspace{0.5cm}
\includegraphics[height=2.5cm]{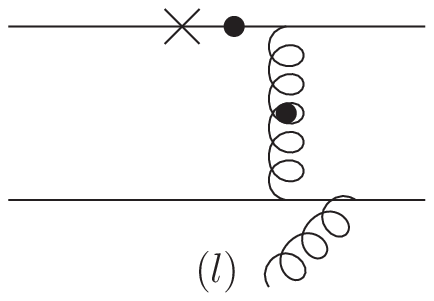}
\caption{Three-parton-to-three-parton diagrams, where the valence
gluon from the initial state is shown, and the possible attachments
of the valence gluon from the final state are represented by dots.}
\label{fig3}
\end{center}
\end{figure}

The other diagrams are classified into several sets, which are
formed by all possible attachments of a hard gluon as displayed in
Figs.~\ref{fig3}(a)-\ref{fig3}(l). Cancellation occurs in each set
of diagrams after applying the Ward identity. Neglecting those
pieces proportional to the parton momenta, we have
\begin{eqnarray}
& &(a)+(b)=(c)+(d)=0,\nonumber\\
& &(e)+(f)+(g)+(h)=0.
\end{eqnarray}
A finite gauge-dependent contribution comes only from
Figs.~\ref{fig3}(i)-\ref{fig3}(l), given by
\begin{eqnarray}
& &(i)+(j)=H^{a\lambda}_{gg}=H^{a\lambda}_{TT}(k_1,k_2+l_2),\nonumber\\
& &(k)+(l)=H^{b\lambda}_{gg}=H^{b\lambda}_{TT}(k_1+l_1,k_2).
\end{eqnarray}

Due to the same expressions of $H^{a\lambda}_{TT}$ and
$H^{a\lambda}_{gT}$, their corresponding matrix elements are
combined into
\begin{eqnarray}
\langle 0|{\bar q}(z)\gamma_5i\gamma^\alpha D_\alpha(0) q(0)
|\pi(P_1)\rangle=0, \label{eq}
\end{eqnarray}
with the equation of motion for the quark field, $\not\!D(0)q(0)=0$.
That is, the gauge invariance holds, when the contributions from
Figs.~\ref{fig1} and \ref{fig2} are combined. The combination of the
matrix elements for $H^{a\lambda}_{Tg}$ and $H^{a\lambda}_{gg}$ also
vanishes according to Eq.~(\ref{eq}). A similar reasoning applies to
the gauge-dependent hard kernels $H^{b\lambda}_{TT}$ and
$H^{b\lambda}_{Tg}$ and to $H^{b\lambda}_{gT}$ and
$H^{b\lambda}_{gg}$: the combination of their matrix elements
vanishes due to the equation of the motion for the quark field in
the final-state pion. This observation completes the proof of the
gauge invariance of the LO three-parton contribution to the pion
electromagnetic form factor at power of $1/Q^2$ in the $k_T$
factorization theorem. The extension of the proof to all orders can
follow the steps outlined in \cite{NL07}. Note that our proof
applies to the collinear factorization theorem too: simply
neglecting transverse momenta in denominators, one can show the
gauge invariance of the three-parton contribution to the pion
electromagnetic form factor in the collinear factorization theorem.

\section{THREE-PARTON CONTRIBUTION}

In this section we calculate the three-parton contribution to the
pion electromagnetic form factor. Start with the gauge-invariant
twist-3 matrix element
\begin{eqnarray}
\langle 0|{\bar q}(z)\sigma^{+}_{\hspace{0.2cm}\alpha'}\gamma_5
iD_\alpha(z') q(0) |\pi(P_1)\rangle,
\label{gim}
\end{eqnarray}
where the subscript $\alpha$ is associated with the vertex the
valence gluon attaches to. The power behavior will not be changed,
and the gauge invariance will not be broken by inserting another
covariant derivative $D^+$. We then exchange $D^+$ and $D_\alpha$,
take the difference of $D^+ D_\alpha$ and $D_\alpha D^+$, and apply
the identity $[D^+,D_\alpha]=-igG^+_{\hspace{0.2cm}\alpha}$. It is
then equivalent to employ the following alternative matrix element
\cite{EKT06}, which defines the three-parton twist-3 pion wave
function $T(z,z')$,
\begin{eqnarray}
\langle 0|{\bar q}(z)\sigma^+_{\hspace{0.2cm}\alpha'}\gamma_5
gG^+_{\hspace{0.2cm}\alpha}(z') q(0) |\pi(P_1)\rangle=if_\pi
m_0(P_1^+)^2g^T_{\alpha\alpha'}T(z,z'),\label{gi3}
\end{eqnarray}
with the chiral scale $m_0=m_\pi^2/(m_u+m_d)$, $m_\pi$, $m_u$, and
$m_d$ being the pion, $u$ quark and $d$ quark masses, respectively.
The operators with other spin structures contribute at higher
twists: for example, the operator $\gamma_\mu\gamma_5
G_{\alpha\beta}$ gives a three-parton twist-4 contribution, and
$\gamma_5 G_{\alpha\beta}$ does not contribute
\cite{BF90,PB3,BBL06}. With Eq.~(\ref{gi3}), one may verify the
gauge invariance of a hard kernel in the $k_T$ factorization theorem
by demonstrating the cancellation between the gauge-dependent
contributions from the operators $\partial^+ A_\alpha$ and
$\partial_\alpha A^+$ \cite{EKT06}.

Below we derive the hard kernels from the
three-parton-to-three-parton diagrams corresponding to
Eq.~(\ref{gi3}) in the Feynman gauge ($\lambda=0$). Choosing this
gauge, the operator $\partial_\alpha A^+$ does not contribute, so
only $\partial^+ A_\alpha$ is relevant. The three momenta
$P_1-k_1-l_1, k_1$, and $l_1$ are assigned to the initial-state
quark, antiquark, and gluon, respectively, and $P_2-k_2-l_2, k_2$,
and $l_2$ to the final-state quark, anti-quark, and gluon,
respectively. We have the structures for the initial- and
final-state pions
\begin{eqnarray}
& &\frac{1}{4}\sigma^{-\alpha'}\gamma_5
\frac{i}{l_1^+}if_\pi m_0 (P_1^+)^2 g^T_{\alpha\alpha'}=
-\frac{i}{4y_1}\not P_1\gamma^T_{\alpha}\gamma_5f_\pi m_0,\nonumber\\
& &-\gamma_5\frac{1}{4}\sigma^{+\beta'} \frac{-i}{l_2^-}(-i)f_\pi
m_0 (P_2^-)^2 g^T_{\beta\beta'}= \frac{i}{4y_2}\gamma_5\not
P_2\gamma^T_{\beta}f_\pi m_0,
\end{eqnarray}
where the gluon momentum fraction $y_2$ is defined by $y_2=l_2^-/P_2^-$,
and the gamma matrix $\gamma^T$ involves only transverse components.

There are totally 196 diagrams for the three-parton-to-three-parton
scattering, which can be divided into four categories\footnote{We
thank the Referee for suggesting this classification of diagrams.}. 
Category A
contains 20 quark-gluon configurations, in which neither of the valence
gluons attaches to the hard gluon line. Each configuration allows 6
different attachments for the photon line. We further divide this
category into two groups as shown in Fig.~\ref{fig6}, where both
valence gluons attach to the same quark or anti-quark line, and in
Fig.~\ref{fig7}, where one valence gluon attaches to the quark line
and another to the anti-quark line. Only the diagrams giving nonvanishing
contributions are displayed. As observed in Appendix B, the
amplitudes with both the valence gluons attaching to the quark line
are power-suppressed. Category B contains 8 quark-gluon
configurations, in which one of the valence gluons attaches to the hard
gluon line. Each configuration allows 5 different attachments for the
photon line, among which those with nonvanishing
contributions are displayed in Fig.~\ref{fig8}. Category C contains
4 quark-gluon configurations, where both the valence gluons are connected
to the hard gluon. Each configuration allows 4 different attachments
for the photon line, among which those with nonvanishing
contributions are displayed in Fig.~\ref{fig9}. Category D contains
4 quark-gluon configurations, where the two valence gluons scatter
via a three-gluon vertex as shown in Fig.~\ref{figC}. Each
configuration allows 5 different attachments for the photon line.
Since this category of diagrams does not contribute, we shall not
discuss them further. Besides, when a valence gluon
attaches to a valence quark, the diagram should be regarded as
being from an effective two-parton Fock state, and will not
be calculated.

We extract the hard kernels proportional to the final-state momentum
$P_{2\mu}$. The hard kernels proportional to $P_{1\mu}$ can be
obtained by exchanging the kinetic variables of the initial- and
final-state pions. Adopting the electric charge $e$, instead of the
quark charge $e_u$ or $e_d$, we have taken into account the diagrams
with the virtual photon attaching to the anti-quark line.
Figure~\ref{fig9}(a) with a four-gluon vertex gives the dominant
three-parton contribution
\begin{eqnarray}
H_{3}& =& ieg^2\frac{f_\pi^2 m_0^2}{16y_1y_2}
\frac{N_c^2}{8(N_c^2-1)} \frac{tr[ \gamma^\sigma \gamma_5\not
P_2\gamma^{T\beta}\gamma_\mu (\not P_1-\not k_2-\not
l_2)\gamma^\nu\not
P_1\gamma^{T\alpha}\gamma_5]}{(P_1-k_2-l_2)^2(k_1-k_2)^2(k_1-k_2+l_1-l_2)^2}
(g_{\alpha \nu} g_{\beta \sigma} +g_{\alpha \sigma} g_{\beta \nu}-2
g_{\alpha\beta} g_{\sigma\nu})\nonumber\\
& = & -i eg^2\frac{f_\pi^2 m_0^2}{16y_1y_2} \frac{ N_c^2 }{N_c^2-1}
\frac{tr[ \not P_2\gamma_\mu \not P_2\not P_1]}{
(P_1-P_2)^2(k_1-k_2+l_1-l_2)^2(k_1-k_2)^2}.\label{aa1}
\end{eqnarray}
To arrive at the second line, we have made an approximation
according to the power counting $Q^2\gg xQ^2, yQ^2\gg k_T^2$
\cite{LWS10}, under which the TMD term in the following denominator
is neglected,
\begin{eqnarray}
\frac{(\not k_2+\not l_2)}{(P_1-k_2-l_2)^2}=
\frac{(k_2+l_2)^-\gamma^+}{-2P_1^+(k_2+l_2)^-}=
\frac{P_2^-\gamma^+}{-2P_1^+P_2^-}=\frac{\not P_2}{(P_1-P_2)^2}.
\end{eqnarray}
The expressions for other three-parton-to-three-parton diagrams are
collected in Appendix B.

Since the Sudakov factor for exclusive QCD processes was derived in
the space of impact parameters \cite{BS,LS}, we Fourier transform
Eq.~(\ref{aa1}). The $k_T$ factorization formula for the pion
electromagnetic pion form factor from Fig.~\ref{fig9}(a) is then
written as
\begin{eqnarray}
F_{3}(Q^2)&=& \pi\alpha_s f_\pi^2 m_0^2\frac{ N_c^2}{N_c^2-1
}\int_0^1 dx_1\int_0^{1-x_1}  \frac{d y_1}{y_1} \int_0^1
dx_2\int_0^{1-x_2}
\frac{d y_2}{y_2}\nonumber\\
& &\times \Phi(x_1,y_1)\Phi(x_2,y_2)
K\left(\sqrt{(x_1+y_1)(x_2+y_2)}Q\right)K(\sqrt{x_1
x_2}Q),\label{f3}
\end{eqnarray}
where the three-parton pion distribution amplitude $\Phi(x_1,y_1)$
corresponds to $T(z,z')$ in Eq.~(\ref{gi3}) in the space of momentum
fractions. The functions $K$, arising from the Fourier
transformation of the TMD denominators $(k_1-k_2+l_1-l_2)^2$ and
$(k_1-k_2)^2$ in Eq.~(\ref{aa1}), are defined by
\begin{eqnarray}
K(t) = \int_0^{1/\Lambda} b db K_0(tb)\exp[-s(P_1^+,b)],\label{kt}
\end{eqnarray}
in which $K_0$ is the modified Bessel function, and the explicit expression
of the Sudakov exponent $s(P_1^+,b)$ is referred to \cite{LS,KLS}.
We have kept only the most effective piece of the Sudakov evolution
in the small $x$ region, that results from the gluon exchanges
between the energetic valence quark and the Wilson line associated
with it. Because the Sudakov factor $\exp[-s(P_1^+,b)]$ diminishes
at $b=1/\Lambda$, with the QCD scale $\Lambda\approx 0.3$ GeV, the
upper bound of the integration variable has been set to $1/\Lambda$
in Eq.~(\ref{kt}). For order-of-magnitude estimate and for
demonstrating the smallness of the three-parton contribution, we do
not consider the renormalization-group evolution from the low scale,
at which $\Phi$ is defined, to the scale of the hard kernel. The
coupling constant is also assumed to be a constant $\alpha_s=0.5$.

\begin{figure}[t]
\begin{center}
\includegraphics[height=15.0cm]{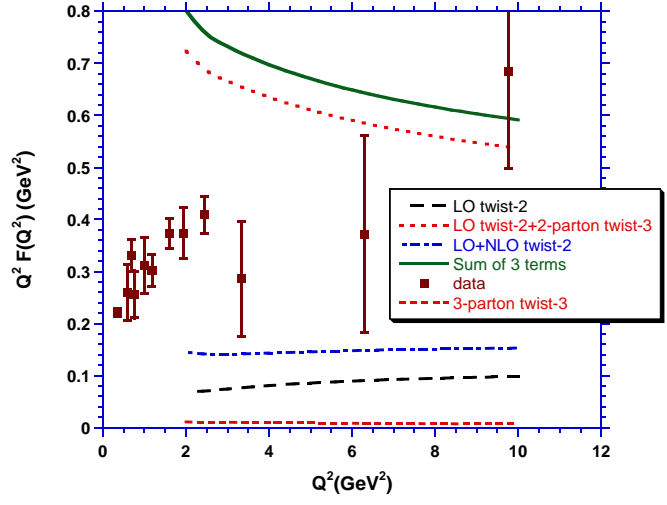}%\hspace{0.5cm}
\caption{$Q^2$ dependence of the three-parton contribution to the
pion electromagnetic form factor. The LO twist-2, NLO twist-2, and
LO two-parton twist-3 contributions are quoted from Fig.~8 in
\cite{LWS10}, which were derived using the non-asymptotic models for
the pion distribution amplitudes.}
\label{figd}
\end{center}
\end{figure}

Fourier transforming the hard kernels in Appendix B into the
impact-parameter space, we construct the corresponding $k_T$
factorization formulas similar to Eq.~(\ref{f3}). We then add all
the contributions to the pion electromagnetic form factor, and
employ the model of the three-parton twist-3 pion distribution
amplitude \cite{BF90,PB3}
\begin{eqnarray}
\Phi(x_1,y_1)=360\eta_3 x_1(1-x_1-y_1)y_1^2
\left[1+\frac{\omega_3}{2}(7y_1-3)\right],
\end{eqnarray}
for a numerical analysis, with the parameters $\eta_3=0.015$ and
$\omega_3=-3$, and $1-x_1-y_1$ being the momentum fraction of the
valence quark. The total three-parton contribution to $Q^2F(Q^2)$,
$F(Q^2)$ being the pion electromagnetic form fact, is displayed in
Fig.~\ref{figd}. The curve exhibits a decrease in $Q^2$ (though not
obvious in the figure) compared to the LO and NLO twist-2
contributions, indicating that this contribution is
power-suppressed. It is only about 5\% of the sum of those evaluated
in \cite{LWS10,CKO09}, including the LO twist-2, NLO twist-2, and LO
two-parton twist-3 pieces. That is, the three-parton contribution is
not crucial for accommodating the experimental data of $Q^2F(Q^2)$
\cite{Huber:2008id,Bebek:1977pe} in the whole accessible range of
$Q^2$ up to 10 GeV$^2$. The only important subleading contribution
to the pion electromagnetic form factor that have been investigated
so far comes from the chirally enhanced two-parton twist-3 one.

\section{CONCLUSION}

In this paper we have applied the $k_T$ factorization theorem to the
study of the power-suppressed three-parton contribution to the pion
electromagnetic form factor. It was demonstrated that the gauge
invariance of the $k_T$-dependent hard kernel holds for this power
correction: the gauge dependence proportional to parton transverse
momenta from the two-parton Fock state and the gauge dependence
associated with the three-parton Fock state cancel each other. We
have calculated the three-parton-to-three-parton hard kernel at LO,
and found that the three-parton contribution is about 5\% of the sum
of the LO twist-2, NLO twist-2, and LO two-parton twist-3 ones in
the whole range of experimentally accessible $Q^2$. Our analysis
shows that the power expansion for this exclusive process might be
reliable in the $k_T$ factorization theorem. At the same power of
$1/Q^2$, the two-parton twist-4 contribution should be taken into
account, which has been studied in the framework of light-cone sum
rules \cite{BK02}. We shall calculate this correction in the $k_T$
factorization theorem in the future, which involves a twist-2
distribution amplitude from one side and a two-parton twist-4
distribution amplitude from the other side. We shall also extend our
framework to exclusive $B$ meson decays, for which three-parton
contributions have been analyzed in light-cone sum rules,
\cite{K01}, in the QCD (collinear) factorization \cite{Yeh08} and in
the soft-collinear effective theory \cite{ARS06}.

We thank Z.T. Wei for his suggestion, which initiates this project.
We also thank V. Braun, S. Brodsky, J.P. Ma and Y.M. Wang for useful
discussions. This work was supported in part by the National Science
Council of R.O.C. under Grant No. NSC-98-2112-M-001-015-MY3, and by
the National Center for Theoretical Sciences.

\appendix

\section{GAUGE DEPENDENCE}

In this appendix we present the detailed derivation of the
gauge-dependent hard kernels from Fig.~\ref{fig2}. The attachment
$A$ contains, with the color factor in Eq.~(\ref{ca}), the
gauge-dependent piece
\begin{eqnarray}
H^{\lambda}_A=ieg^2\lambda\frac{1}{32N_c^2} \frac{tr[(\not k_1-\not
k_2) \gamma_5\gamma_{\beta}\gamma_\mu (\not P_1-\not
k_2)\gamma^\alpha(\not P_1-\not l_1-\not k_2)(\not k_1-\not k_2)
\gamma_{\alpha}\gamma_5]}{(P_1-k_2)^2(P_1-l_1-k_2)^2(k_1-k_2)^4}.\label{at1}
\end{eqnarray}
Inserting the identity $\not k_1-\not k_2=(\not P_1-\not l_1-\not
k_2)-(\not P_1-\not l_1-\not k_1)$ into Eq.~(\ref{at1}), we obtain
\begin{eqnarray}
H^{\lambda}_A&=&ieg^2 \lambda\frac{1}{32N_c^2}\Bigg\{\frac{tr[(\not
k_1-\not k_2) \gamma_5\gamma_{\beta}\gamma_\mu (\not
P_1-\not k_2)\gamma^\alpha
\gamma_{\alpha}\gamma_5]}{(P_1-k_2)^2(k_1-k_2)^4}\nonumber\\
& &-\frac{tr[(\not k_1-\not k_2) \gamma_5
\gamma_{\beta}\gamma_\mu (\not P_1-\not k_2)\gamma^\alpha(\not
P_1-\not l_1-\not k_2)(\not P_1-\not k_1-\not l_1)
\gamma_{\alpha}\gamma_5]}
{(P_1-k_2)^2(P_1-l_1-k_2)^2(k_1-k_2)^4}\Bigg\}.\label{13}
\end{eqnarray}
The second term, proportional to the momentum $P_1- k_1- l_1$ of the
incoming valence quark, should be dropped, since there is the
valence gluon $A^\alpha$ from the initial state already. Taking the
derivative of the first term with respect to $k_{2\beta}$, and then
substituting $k_1= x_1P_1$ and $k_2= x_2P_2$ into the numerator, we
have Eq.~(\ref{aA}).

The diagram with the attachment $B$ of the valence gluon to the hard
gluon line produces the gauge-dependent hard kernel
\begin{eqnarray}
H^{\lambda}_B&=&-eg^2 \frac{1}{8N_c}\frac{tr[
\gamma_{\delta'}\gamma_5\gamma_{\beta}\gamma_\mu (\not P_1-\not
k_2)\gamma_{\nu'}\gamma^{\alpha}\gamma_5]}{(P_1-k_2)^2}
tr(T^dT^bT^c)\Gamma^{dba}_{\delta\nu\alpha}\nonumber\\
& &\times\Bigg[\lambda\frac{g^{\delta\delta'}}{(k_1-k_2)^2}
\frac{(k_1-k_2+l_1)^\nu(k_1-k_2+l_1)^{\nu'}}{(k_1-k_2+l_1)^4}+\lambda
\frac{(k_1-k_2)^\delta(k_1-k_2)^{\delta'}}{(k_1-k_2)^4}
\frac{g^{\nu\nu'}}{(k_1-k_2+l_1)^2}\nonumber\\
& &-\lambda^2\frac{(k_1-k_2)^\delta(k_1-k_2)^{\delta'}}{(k_1-k_2)^4}
\frac{(k_1-k_2+l_1)^\nu(k_1-k_2+l_1)^{\nu'}}{(k_1-k_2+l_1)^4}\Bigg],
\end{eqnarray}
with the triple-gluon vertex,
\begin{eqnarray}
\Gamma^{dba}_{\delta\nu\alpha}=f^{dba}
[g_{\alpha\nu}(2l_1+k_1-k_2)_\delta
+g_{\nu\delta}(2k_2-2k_1-l_1)_\alpha
+g_{\delta\alpha}(k_1-k_2-l_1)_\nu], \label{tg3}
\end{eqnarray}
$f^{dba}$ being a antisymmetric tensor. Using the identity
\begin{eqnarray}
tr(T^dT^bT^c)=\frac{1}{4}(d^{dbc}+if^{dbc}),\;\;\;
d^{dbc}f^{dba}=0,\;\;\;f^{dbc}f^{dba}=N_c\delta^{ac},
\end{eqnarray}
$d^{dbc}$ being a symmetric tensor, the above amplitude becomes
\begin{eqnarray}
H^{\lambda}_B&=&-ieg^2 \frac{1}{32}\frac{tr[
\gamma_{\delta'}\gamma_5\gamma_{\beta}\gamma_\mu (\not P_1-\not
k_2)\gamma_{\nu'}\gamma^{\alpha}\gamma_5]}{(P_1-k_2)^2}
\nonumber\\
& &\times [g_{\alpha\nu}(2l_1+k_1-k_2)_\delta
+g_{\nu\delta}(2k_2-2k_1-l_1)_\alpha
+g_{\delta\alpha}(k_1-k_2-l_1)_\nu]\nonumber\\
& &\times\Bigg[\lambda\frac{g^{\delta\delta'}}{(k_1-k_2)^2}
\frac{(k_1-k_2+l_1)^\nu(k_1-k_2+l_1)^{\nu'}}{(k_1-k_2+l_1)^4}+\lambda
\frac{(k_1-k_2)^\delta(k_1-k_2)^{\delta'}}{(k_1-k_2)^4}
\frac{g^{\nu\nu'}}{(k_1-k_2+l_1)^2}\nonumber\\
& &-\lambda^2\frac{(k_1-k_2)^\delta(k_1-k_2)^{\delta'}}{(k_1-k_2)^4}
\frac{(k_1-k_2+l_1)^\nu(k_1-k_2+l_1)^{\nu'}}{(k_1-k_2+l_1)^4}\Bigg].
\label{hB}
\end{eqnarray}

The $\lambda^2$ term leads to
\begin{eqnarray}
& &ieg^2 \frac{1}{32}\lambda^2\frac{tr[(\not k_1-\not
k_2)\gamma_5\gamma_{\beta}\gamma_\mu (\not P_1-\not k_2)(\not
k_1-\not k_2+\not l_1)\gamma_{\alpha}\gamma_5]}{(P_1-k_2)^2(k_1-k_2)^4(k_1-k_2+l_1)^4}\nonumber\\
& &\times\Big[(k_1-k_2)\cdot(2l_1+k_1-k_2)(k_1-k_2+l_1)^\alpha+
(k_1-k_2)\cdot(k_1-k_2+l_1)(2k_2-2k_1-l_1)^\alpha\nonumber\\
& &+(k_1-k_2-l_1)\cdot(k_1-k_2+l_1)(k_1-k_2)^\alpha\Big].
\end{eqnarray}
Inserting the identity $\not k_1-\not k_2+\not l_1=(\not P_1-\not
k_2)-(\not P_1-\not k_1-\not l_1)$, it is easy to see that the first
term of the identity gives an expression which is cancelled by the corresponding
one from the attachment $G$. The second term, being proportional to
the momentum $P_1- k_1- l_1$ of the incoming valence quark, should
be neglected.

We then consider the $\lambda$ terms. The first $\lambda$ term in
Eq.~(\ref{hB}) gives
\begin{eqnarray}
-ieg^2 \frac{1}{32}\lambda\frac{tr[(2\not l_1+\not k_1-\not
k_2)\gamma_5\gamma_{\beta}\gamma_\mu (\not P_1-\not k_2)(\not
k_1-\not k_2+\not l_1)\gamma_{\alpha}\gamma_5]}
{(P_1-k_2)^2(k_1-k_2)^2(k_1-k_2+l_1)^4}(k_1-k_2+l_1)^\alpha
\nonumber\\
-ieg^2 \frac{1}{32}\lambda\frac{tr[(\not k_1-\not k_2+\not
l_1)\gamma_5\gamma_{\beta}\gamma_\mu (\not P_1-\not k_2)(\not
k_1-\not k_2+\not l_1)\gamma_{\alpha}\gamma_5]}
{(P_1-k_2)^2(k_1-k_2)^2(k_1-k_2+l_1)^4}(2k_2-2k_1-l_1)^\alpha
\nonumber\\
-ieg^2 \frac{1}{32}\lambda\frac{tr[\gamma^\alpha\gamma_5
\gamma_{\beta}\gamma_\mu (\not P_1-\not k_2)(\not k_1-\not k_2+\not
l_1)\gamma_{\alpha}\gamma_5]}
{(P_1-k_2)^2(k_1-k_2)^2(k_1-k_2+l_1)^4}(k_1-k_2+l_1)\cdot(k_1-k_2-l_1),
\end{eqnarray}
which are all negligible for the same reason as for the
$\lambda^2$ term. Hence, $H^{\lambda}_B$ receives a contribution
only from the second $\lambda$ term,
\begin{eqnarray}
H^{\lambda}_B&=&-ieg^2 \frac{1}{32}\lambda\frac{tr[(\not k_1-\not
k_2)\gamma_5\gamma_{\beta}\gamma_\mu (\not P_1-\not
k_2)\gamma^\alpha\gamma_{\alpha}\gamma_5]}
{(P_1-k_2)^2(k_1-k_2)^4(k_1-k_2+l_1)^2}
(k_1-k_2)\cdot(2l_1+k_1-k_2)\nonumber\\
&=&-ieg^2 \frac{1}{32}\lambda\frac{tr[(\not k_1-\not k_2)\gamma_5
\gamma_{\beta}\gamma_\mu (\not P_1-\not k_2)\gamma^\alpha
\gamma_{\alpha}\gamma_5]}{(P_1-k_2)^2(k_1-k_2)^4}.
\end{eqnarray}
To arrive at the second line, the higher-power term $l_1^2=-l_{1T}^2$
has been added, so we have $(k_1-k_2)\cdot(2l_1+k_1-k_2)+l_1^2=(k_1-k_2+l_1)^2$.
The differentiation of the above expression with respect to
$k_{2\beta}$ leads to Eq.~(\ref{B}).
The diagrams with other attachments in Fig.~\ref{fig2} can be
calculated in a similar way, so the detail will not be presented
here.

\section{HARD KERNELS}

\begin{figure}[t]
\begin{center}
\includegraphics[height=3.2cm]{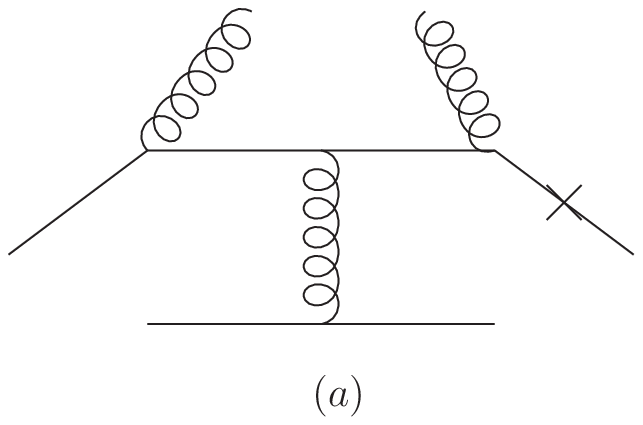}\hspace{0.5cm}
\includegraphics[height=3.2cm]{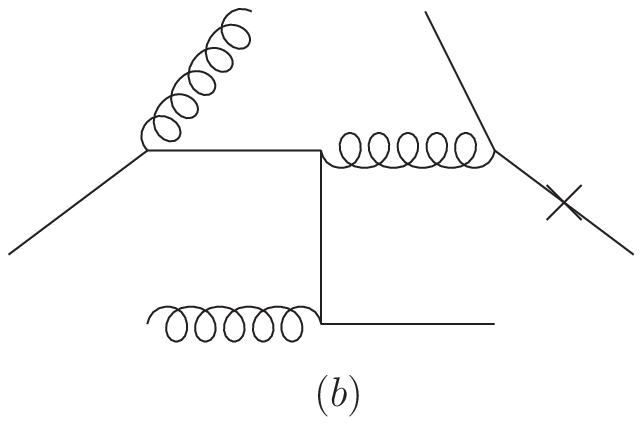}\hspace{0.5cm}
\includegraphics[height=3.2cm]{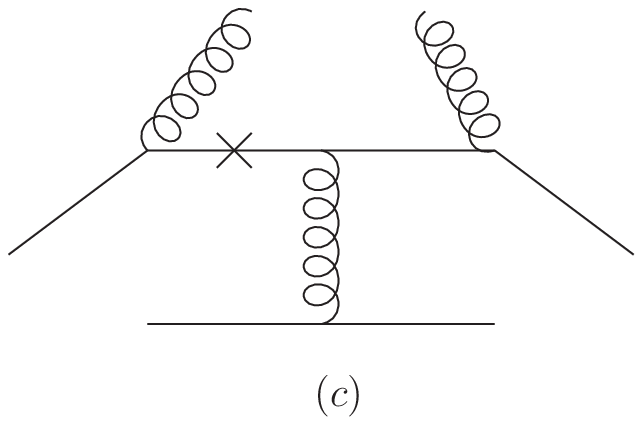}
\caption{Three-parton-to-three-parton diagrams in Category A, where both valence
gluons attach to the same quark or anti-quark line.} \label{fig6}
\end{center}
\end{figure}

\begin{figure}[t]
\begin{center}
\includegraphics[height=4.0cm]{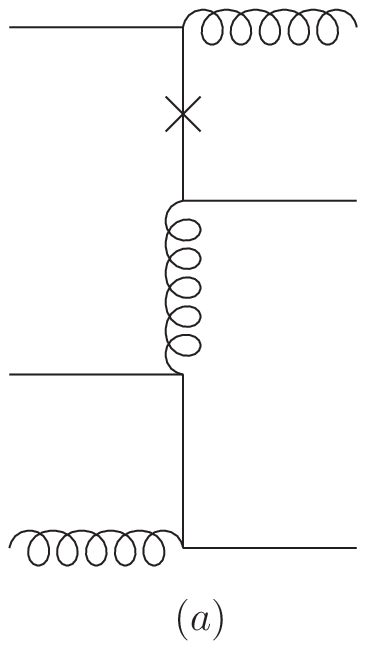}\hspace{1.5cm}
\includegraphics[height=4.0cm]{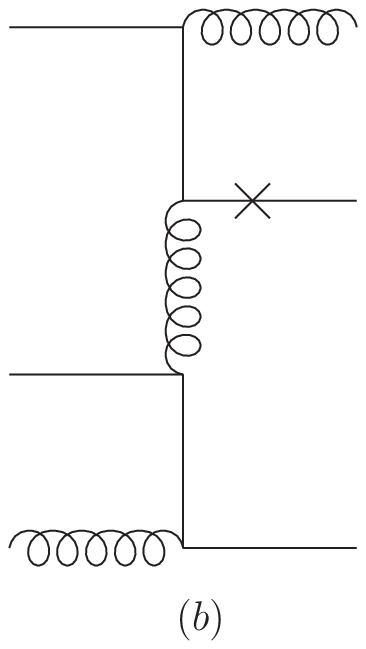}\hspace{1.5cm}
\includegraphics[height=4.0cm]{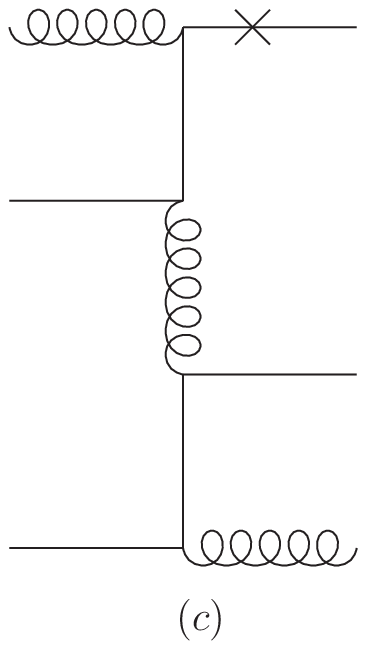}
\caption{Three-parton-to-three-parton diagrams in Category A, where one
valence gluon attaches to the quark line and another to the
anti-quark line.}
\label{fig7}
\end{center}
\end{figure}

In this appendix we collect the expressions of the
three-parton-to-three-parton hard kernels for the pion
electromagnetic form factor in the Feynman gauge. Start with Category A
defined in Sec.~III. When the attachments of
the two valence gluons are arranged in the way that the hard gluon
vertices sandwich the spin structures associated with the pions, the
contribution diminishes because of $\gamma^{\nu} \not\!
P_1\gamma^T_{\alpha} \gamma_{\nu}=0$ or $\gamma^{\nu} \not\!
P_2\gamma^T_{\beta} \gamma_{\nu}=0$. The nonvanishing amplitudes
come from Figs.~\ref{fig6} and \ref{fig7}, which are written as
\begin{eqnarray}
H_{6a}&=& -ieg^2\frac{N_c^2+1}{N^2_c(N_c^2-1)}\frac{tr(\not
P_2 \gamma_\mu \not P_2 \not P_1 \not P_2\not P_1
)}{(P_1-P_2)^2(P_1-k_2-l_2)^2(P_1-k_1-l_1-P_2)^2(k_1-k_2)^2},\\
H_{6b}&=& ieg^2\frac{N_c^2+1}{N^2_c(N_c^2-1)} \frac{tr( \not
P_2 \gamma_\mu \not P_2\not P_1 \not l_2\not l_1
)}{(P_1-P_2)^2(k_1-k_2+l_1-l_2)^2(k_1 -l_2 )^2(k_2-l_1)^2},\label{7b}\\
H_{6c}&=& ieg^2\frac{N_c^2+1}{N^2_c(N_c^2-1)} \frac{tr(\not
P_2\not P_1 \not P_2\gamma_\mu \not P_2\not P_1
)}{(P_2-k_2-l_2-P_1)^2(P_2-k_1-l_1)^2(P_1-k_1 - l_1 -P_2
)^2(k_1-k_2)^2},\\
H_{7a}&=& ieg^2\frac{1}{N^2_c (N_c^2-1)}\frac{tr(\not
P_2\not l_1\not P_2\gamma_\mu \not P_2 \not P_1
)}{(P_1-k_1-l_1-P_2)^2(k_1-k_2+l_1)^2(l_1-k_2)^2(P_2-k_1-l_1)^2}, \\
H_{7b}&=& -i eg^2\frac{1}{N^2_c (N_c^2-1)}\frac{tr(\not
P_2\gamma_\mu\not P_2\not P_1\not P_2\not l_1
)}{(P_1-P_2)^2(k_1-k_2+l_1)^2(l_1-k_2)^2(P_1-k_1-l_1-P_2)^2},\\
H_{7c}&=&ieg^2\frac{1}{N^2_c (N_c^2-1)}\frac{tr( \not
P_2\gamma_\mu \not P_2\not P_1\not l_2\not P_1
)}{(P_1-P_2)^2(k_1-k_2)^2(k_1-l_2)(P_1-k_2-l_2)^2}.
\end{eqnarray}
It is observed that the results from Figs.~\ref{fig6}(a) and
\ref{fig6}(c) are suppressed by a power of $1/Q$ compared to
Fig.~\ref{fig6}(b). That is, when both the valence gluons attach to
the quark line, the contribution is power-suppressed. The hard kernels
from Fig.~\ref{fig7} are of the same power as Eq.~(\ref{7b}).

\begin{figure}[t]
\begin{center}
\includegraphics[height=4.0cm]{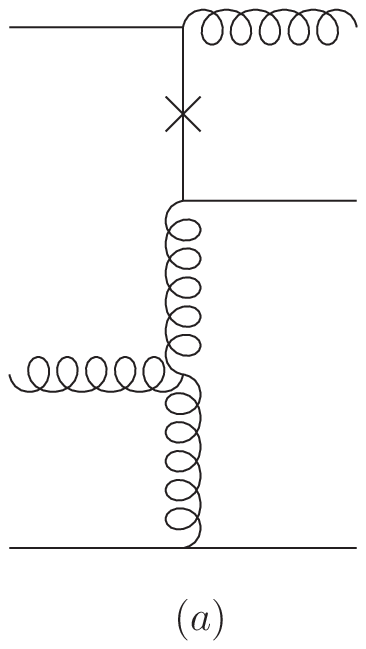}\hspace{1.5cm}
\includegraphics[height=4.0cm]{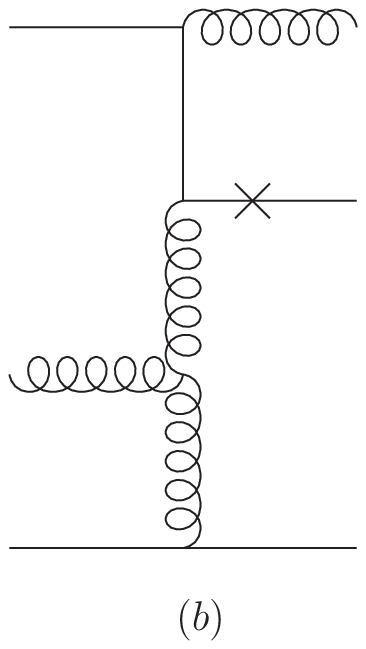}\hspace{1.5cm}
\includegraphics[height=4.0cm]{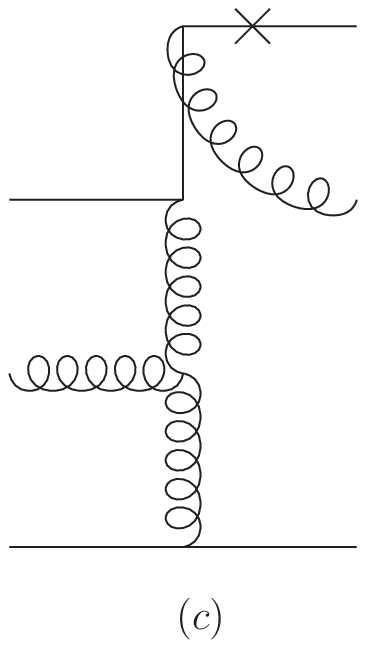}\\
\includegraphics[height=4.0cm]{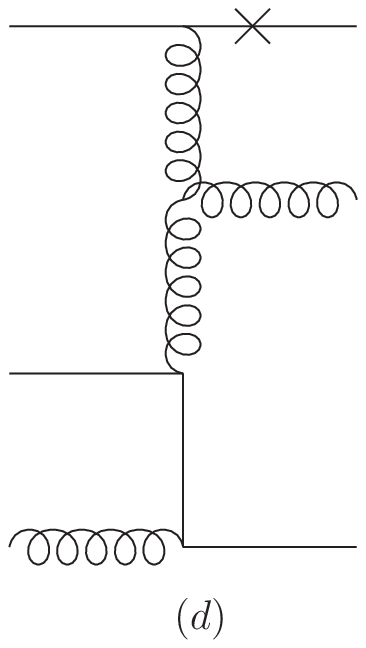}\hspace{1.5cm}
\includegraphics[height=4.0cm]{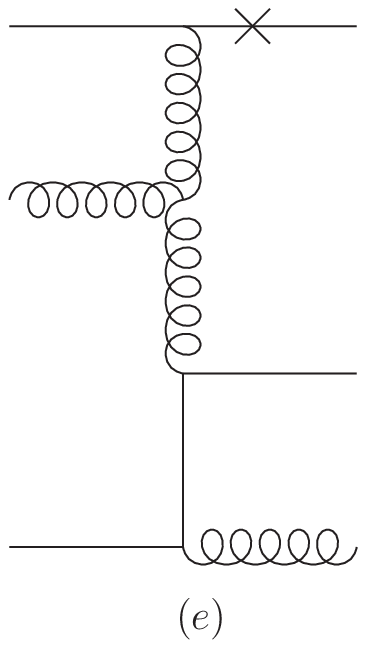}\hspace{1.5cm}
\includegraphics[height=4.0cm]{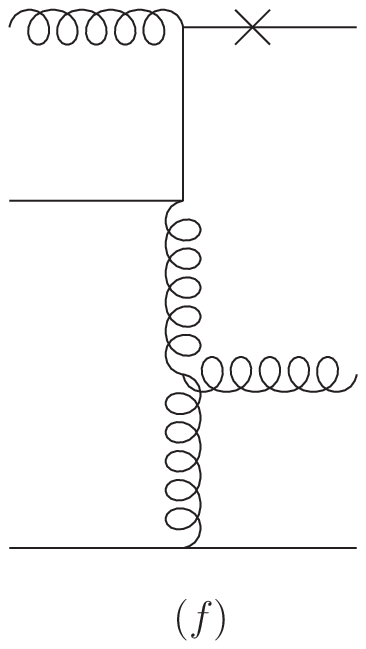}
\caption{Three-parton-to-three-parton diagrams in Category B, where one valence
gluon attaches to the hard gluon line.} \label{fig8}
\end{center}
\end{figure}

Figure~\ref{fig8} from Category B gives the hard kernels
\begin{eqnarray}
H_{8a}&=& -\frac{i}{2}eg^2 \frac{ 1}{N_c^2-1}\frac{tr[\not
P_2(\not k_1 -\not l_1) \not P_2\gamma_\mu \not P_2\not P_1
]}{(P_2-k_1-l_1)^2(k_1-k_2+l_1)^2(k_1-k_2)^2(P_1-k_1-l_1-P_2)^2}, \\
H_{8b}&=& \frac{i}{2}eg^2\frac{1}{N_c^2-1}\frac{tr[\not
P_2\gamma_\mu \not P_2 (\not k_1 - \not l_1)\not P_2\not
P_1]}{(P_1-P_2)^2(k_1-k_2+l_1)^2(k_1-k_2)^2
(P_1-k_1-l_1-P_2)^2}, \\
H_{8c}&=& \frac{i}{2}eg^2\frac{tr( \not P_2\gamma_\mu\not
P_2\not P_1
)}{(P_1-P_2)^2(k_1-k_2+l_1)^2(k_1-k_2)^2}, \\
H_{8d}&=& -\frac{i}{2}eg^2 \frac{1}{N_c^2-1}\frac{tr[\not
P_2\gamma_\mu\not P_2\not P_1(\not k_2+2\not l_2)\not l_1
]}{(P_1-P_2)^2(k_1-k_2+l_1)^2(l_1-k_2)^2(k_1-k_2+l_1-l_2)^2},
\\
H_{8e}&=& -\frac{i}{2}eg^2\frac{1}{N_c^2-1}\frac{tr[\not
P_2\gamma_\mu\not P_2 \not P_1\not l_2(\not k_1+2\not l_1) ]}
{(P_1-P_2)^2(k_1-k_2+l_1-l_2)^2(k_1-l_2)^2 (k_1-k_2-l_2)^2}, \\
H_{8f}&=& -\frac{i}{2}eg^2\frac{1}{N_c^2-1}\frac{tr[ \not
P_2 \gamma_\mu\not P_2\not P_1(\not k_2 - \not l_2)\not P_1
]}{(P_1-P_2)^2(k_1-k_2)^2 (k_1-k_2-l_2)^2(P_1-k_2-l_2)^2}.
\end{eqnarray}

\begin{figure}[t]
\begin{center}
\includegraphics[height=4.0cm]{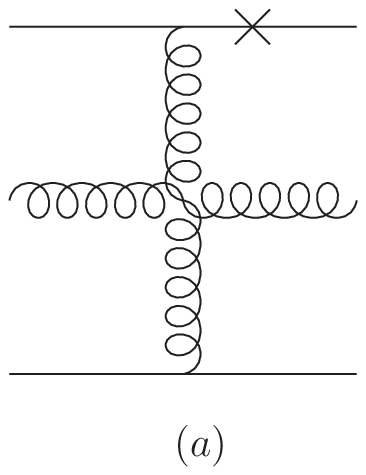}\hspace{1.5cm}
\includegraphics[height=4.0cm]{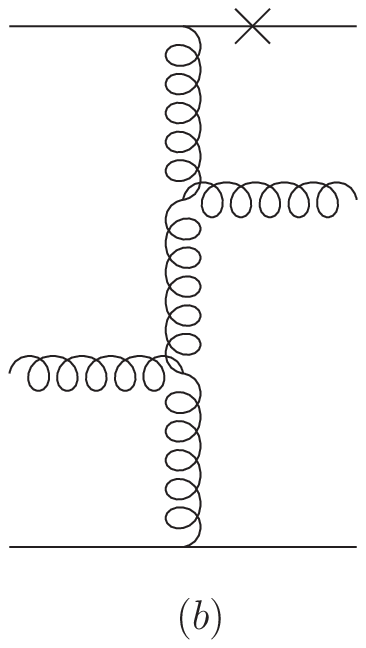}\hspace{1.5cm}
\includegraphics[height=4.0cm]{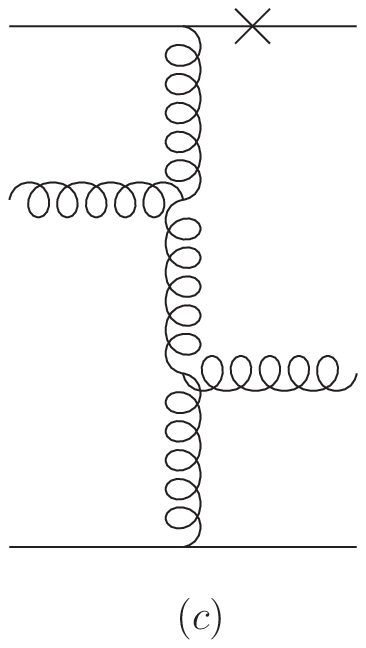}
\caption{Three-parton-to-three-parton diagrams in Category C, where both valence
gluons attach to the hard gluon line.} \label{fig9}
\end{center}
\end{figure}

Figure~\ref{fig9} from Category C contributes the hard kernels
\begin{eqnarray}
H_{9a} & = & -i eg^2 \frac{ N_c^2 }{N_c^2-1} \frac{tr( \not
P_2\gamma_\mu \not P_2\not P_1)}{
(P_1-P_2)^2(k_1-k_2+l_1-l_2)^2(k_1-k_2)^2},\\
H_{9b}&=& - \frac{i}{2}eg^2 \frac{N_c^2}{N_c^2-1}\frac{( k_1-
k_2- 2 l_2+l_1) \cdot (k_1 - k_2 - l_1 )   tr(\not P_2 \gamma_\mu
\not P_2\not P_1
)}{(P_1-P_2)^2(k_1-k_2+l_1)^2(k_1-k_2)^2(k_1-k_2+l_1-l_2)^2},\\
H_{9c}&=& -\frac{i}{2}eg^2\frac{N_c^2}{N_c^2-1}\frac{( k_1-
k_2+ l_2) \cdot (k_1 - k_2 + 2 l_1 -l_2)  tr( \not P_2
\gamma_\mu\not P_2 \not P_1 )}{(P_1-P_2)^2(k_1-k_2+l_1-l_2)^2
(k_1-k_2-l_2)^2(k_1-k_2)^2}.
\end{eqnarray}

%%%%%%%%%%%%%%%%%%%%%%%%%%%%%%%%%%%%555555

\end{document}